\documentclass[12pt]{article}
\pdfoutput=1
\usepackage{mathtext}
\usepackage{graphicx}
\usepackage[english]{babel}
\usepackage[utf8]{inputenc}
\usepackage{amsfonts}
\usepackage{amssymb}
\usepackage{amsmath}
\renewcommand{\d}{\delta}
\renewcommand{\b}{\beta}
\newcommand{\g}{\gamma}

\newcommand{\ar}{\longrightarrow}
\newcommand{\w}{\omega}
\newcommand{\s}{\sigma}
\newcommand{\la}{\lambda}
\renewcommand{\a}{\alpha}
\begin{document}
\title{Quantum revivals of non Rabi type in Jaynes-Cummings model}
\author{ Yu.Ozhigov\thanks{ozhigov@cs.msu.su}, \ \ N.Victorova\thanks{nbvictorova@list.ru},\\
Moscow State Lomonosov University, VMK, \\
Department of supercomputers and quantum informatics,\\
Russian State University for the humanities \\ 
Department of Fundamental and Applied Mathematics}

\maketitle

\begin{abstract}
 We studied full revivals of quantum states in the Jaynes-Cannings model.
It is proved that in the case of zero detuning in subspaces generated by two adjacent pairs of energy levels, full revival of the subspace does not exist for any values of the parameters. In the case of non-zero detuning on the contrary, the set of parameters that allows full revival of such subspaces is dense as the subset of all parameters. The nature of these revivals differs from Rabi oscillations in subspaces of a single pair of  energy. In more complex subspaces  the presence of full revival is reduced to particular cases  of 10-th Hilbert problem for rational solutions of systems of nonlinear algebraic equations, which has no algorithmic solution in general case.
 \end{abstract}

\section{Introduction}

Finite-dimensional, so-called qubit models is particularly important for building elements of a quantum computer. It is known, for example, that CNOT gate can be build by nonlinear phase shifts and beam splitters. Moreover, the universal quantum computer theoretically can be built on the optical elements with the obligatory presence of non-linear ones (see. \cite{KLM}, \cite{DF}, further bibliography is in the \cite{KM}). If we ignore the decoherence factor, the fast quantum algorithms that outperform any classical can work on optical elements. Therefore, it is impossible to simulate in conventional terms the dynamics of the optical devices.

This emphasizes the importance of a partial description of the dynamics of quantum optical devices, especially of non-linear type. The main theoretical approach here is Jaynes-Cummings model - JC (\cite{JC}), serving as the basis in studies of non-linear elements. Despite the fact that this model is based on RWA approximation, the processes of long duration can be modelled in the framework of JC. The simplest example is Rabi oscillations for the subspace of the close energies; the more general case is quantum revivals (see, e.g., \cite{Ka}).

The aim of our work is to study quantum revivals that are not reducible to Rabi oscillations (see. e.g., \cite{HZ}). The ability to implement scalable quantum computations on the optical devices is based on the lack of such a revival in the case of resonance: the coincidence of atomic and cavity frequencies. But if there is even a small non-zero detuning between these frequencies, such revivals occur in the subspaces generated by the eigenvectors of different pairs of close energies. This can create an unexpected obstacles in the work of quantum gates with non-linear elements such as a phase shift, which in terms of Fock states has the following form: $|0\rangle\ar |0\rangle,\ |1\rangle\ar |1\rangle,\ |2\rangle\ar -|2\rangle$. The  nature of these obstacles is associated with parasitic resonances that arise due to the commensurability of energy gaps.

Revivals and partial (approximate or for certain types of states) revivals provide a simple and effective tool for studying the unitary dynamics of quantum ensembles (\cite {HA}, \cite {SF}, \cite {RH}). The revival of individual state $|\Psi\rangle$ in the unitary evolution \newline $U_t:\ | \Psi (0) \rangle \ar \ | \Psi (t) \rangle $ in time $t_0 $ means the existence of a real $ \phi $, for which $ | \Psi (t_0) \rangle = e^{i \phi} | \Psi (0) \rangle $, or, in terms of density matrices,
\begin {equation}
\rho (0) = \rho (t_0).
\label{rev}
\end {equation}
  Full revival of the subspace means the existence of a time interval $ t_0 $, such that equality (\ref{rev}) is true for any state $ \rho (0)$ in this subspace  (it is enough to require it for any pure state). We take Plank constant $h =1 $ . Let $ |\psi_j \rangle$ denote the eigenvectors of the Hamiltonian $ H $, with eigenvalues $E_j$,  $j=0,1,\ldots,N-1$, then the evolution of any pure state has the form
\begin {equation}
| \psi (t) \rangle = \sum \limits_j \la_je ^ {- iE_jt} | \psi_j \rangle
\label{ev}
\end {equation}

and the revival of the state $ |\psi (0)\rangle $ is completely determined by the part of the spectrum of $ E_ {i_0}, E_ {i_1}, \ldots, E_ {i_ {l}} $, which corresponds to the minimum subspace  generated by the eigenvectors $ |\psi_j \rangle $, which contains $ |\psi \rangle $. Suppose, for simplicity, that the subspace is generated by the vectors $ |\psi_j \rangle $, such that $ j \in \{0,1,\ldots, N-1\} $. Then the condition of the full revival of this subspace is equivalent to the rationality of all numbers $ r_j = (E_j-E_0) /(E_1-E_0) $ for $ j \in \{ 2, \ldots, N-1 \} $.

Let $ r_j = n_j /d_j $ be irreducible fraction, and $K_1$ is the least common multiple of $d_j, \ j = 1,2, \ldots, N-1 $. Then the time of full revival will be $t = 2 \pi /\d$, where $\d = (E_1-E_0) /K_1$. Therefore practically important parameter - the time of full revival depends on the rationality of  quotients of the gaps between the energy levels.

\section{Qubit model of JC type}

We will work with the qubit model, which basis vectors have the form  \newline $|p_{n-1},p_{n-2},\ldots,p_0, \a\rangle$, where $\sum\limits_{j=0}^{n-1}p_j2^j$ is the number of photons , $\a$ denotes the excitation of the atom ($0$ - ground state, $1$ - excited state).
 Hamiltonian  of the model JC has the form  $H_{JC}=\frac{h\w_a}{2}\s^+\s^-+\frac{h\w_c}{2}a^+a
+h\w_a\g \s^+a_j+h.c.$  In matrix form, with the lexicographical ordering of the basis vectors, the Hamiltonian is block-diagonal matrix, where on the main diagonal stands $0$ at first, and then blocks of the form  
$$\left(
\begin{array}{lll}
& n\w_a+(n-1)\Delta &\ \ \sqrt{n}y\\
&\ \ \sqrt{n}y& n(\w_a+\Delta )
\end{array}
\right)
$$
where $\w_a$ is the frequency of the atom, $\w_c=\w_a+\Delta$ is  the frequency of the cavity, $\Delta\ll \w_a$ is small detuning, $y=\w_a\g$, where $\g$ is the matrix element of dipole momentum of the passage from ground state of atom to the excited state, $n$ denotes the number of the box. Each box thus corresponds to the two dimensional subspace, which we denote ${\cal H}_n$. Full revivals of this subspace take place for every value of $\w_a,\Delta,y$, there are well known Rabi oscillations (see \cite{Sho}). 

We  now consider a pair of subspaces  ${\cal H}_{n-1},\ {\cal H}_n$. We intend to find such parameters  $\w_a,\ \d ,\ y,\ \Delta$, for which there is a full revival of the linear span   ${\cal H}={\cal L}({\cal H}_{n-1},\ {\cal H}_n)$  of these subspaces. This is not the revival of Rabi type, but the nontrivial resonance effect between different pairs of energy levels.

Spectrum of the Hamiltonian $H_{JCH}$, restricted to the subspace  ${\cal H}$, is the set consisting of two  pairs of real numbers

\begin{equation}
\label{Spectrum}
\begin{array}{ll}
&S_{n-1} = \{ \frac{1}{2}(2\w_a+\Delta+2(n-1)(\w_a+\Delta )\pm\sqrt{\Delta^2+4ny^2})\},\\
&S_n = \{ \frac{1}{2}(2\w_a+\Delta+2n(\w_a+\Delta )\pm\sqrt{\Delta^2+4(n+1)y^2})\},
\end{array}
\end{equation}

We introduce the notations $\a=\frac{\Delta}{y},\ \b=\frac{\w_a}{y}$. Then, taking fractions of  energy gaps, as described above, we find that the full revival of the subspace ${\cal H}$ is equivalent to the rationality of two numbers 
\begin{equation}
\frac{2\a +2\b \pm \sqrt{\a^2+4(n+1)}}{2\sqrt{\a^2+4n}}
\label{frac}
\end{equation}
In case of the resunance  ($\a=0$),  subtracting these fractions we come to the rationality of the fraction $\sqrt\frac{n+1}{n}$, that is impossible.
 
We now consider the non-resonance case, when $\a\neq 0$.  For rationality of (\ref{frac}) it is sufficient that the both roots $\sqrt{\a^2+4n)} $   and    $\sqrt{\a^2+4(n+1)}$ are rational, because choosing $\b$, we can guarantee rationality of the both fractions.     

Rationality  of these roots  is ensured by the presence of rational roots of the equation 
 $X^2-Y^2=1$. One of this roots is integer: $X=1,Y=0$. To search for a dense set of roots we present them as the intersection of the  target hyperbola with the  line of the form $(X-1)= tY$, passing through the integer solution found. Expressing variables via the parameter $t$, which we assumed to be rational, we obtain a dense set of solutions:    
\begin{equation}
\label{solution}
X=1+\frac{2t^2}{1-t^2},\ Y=\frac{2t}{1-t^2}.
\end{equation}

Since Jaynes-Cummings model itself is an approximation (about the revival with the members of the counter-rotating terms, i.e. without rotating wave approximation, see, for example, \cite{ZQ}),  the time of  full revival  $T$ is significant. Since the physical value $y$ is proportional to $\w_a$, we can conclude that roughly $T$ is inversely proportional to the number $n$, designating a highest energy level  of considered pairs of subspaces. More subtle dependence $T$ from our parameters $ \w_a, \ \Delta, \ y$  is complex, because $ T $ will increase with the growth of the least common multiple of the denominators of the irreducible representations of $ X $ and $ Y $  
(\ref{solution}). The density distribution of this quantity $LCM (Denom (X), Denom (Y)) $ behaves quite erratically (see Figure 1).
 \begin{figure}
\includegraphics[scale=0.6,
bb=10 -100 200 200]{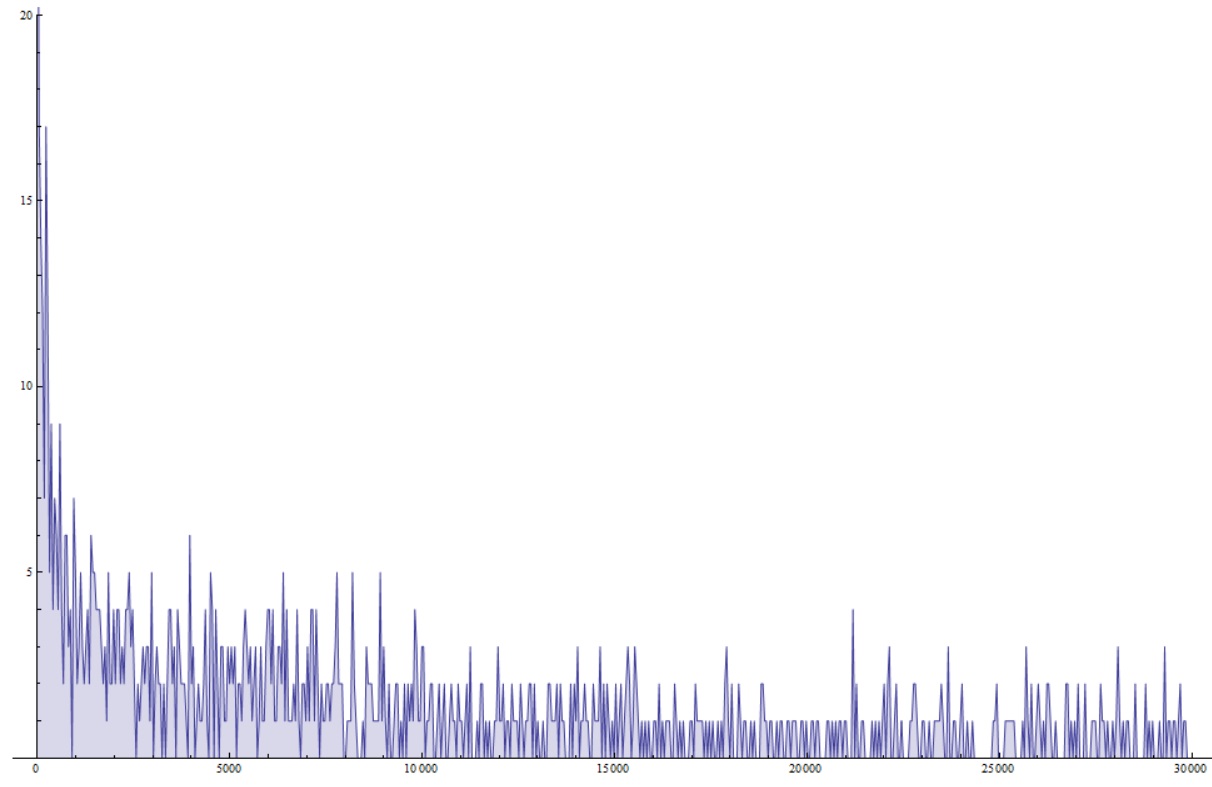}
 \caption{Distribution density values  LCM(Denom(X),Denom(Y)). Calculation is done by scanning the field of t=nd  for d=1/10000, n=1,2,...,30000.}  
\end{figure}

For a couple of energy level pairs that are not adjacent, the revival in the linear span of the subspaces is more sophisticated. Here it is sufficient to find rational solutions  of the equation $ X ^ 2-Y ^ 2 = K $, where $K$ is the distance between pairs of energy levels. Such solutions are known to exist, if $K = k^2 $ for some integer $ k $, or for values $ K $, which are  junior members of the Pythagorean triples, that is represented in the form $K = k_1^2- k_2^2 $ for the integer $k_1, k_2$.

  For a set of $s$ pairs of energy levels located at distances $ K_j $ full revival of the linear span can be reduced to systems of equations in rational numbers of the form $ X_ {j-1}^ 2-X_j^ 2 = K_j, \ j = 1,2, \ldots, s-1 $.  For $ s = 3 $ we can specify a number of partial solutions for such numbers $ K_j $, which are part of Pythagorean triples. For example, a solution of $ X^2-Y^2 = K^2, \ Y^2-Z^2 = L^2 $ can give such $ Y $, which are part of Pythagorean triples, as a highest member, and as a junior as well; examples of such numbers: $ Y = 15,20,30,40 $. This suggests a rich set of revivals for the three pairs of  energy levels simultaneously. On the other hand, for some $ K_j $ this task may not have a solution. In general, for a non-zero detuning, the problem of finding the full revivals is reduced to finding rational solutions of systems of algebraic equations. This problem is equivalent to the solution of Diophantine equations (10-th Hilbert problem), for which there is no algorithmic solution (see. \cite{Ma}).

\section{Conclusion}

We have established the presence of  revivals of non Rabi  type for Jaynes-Cummings model in subspaces generated by adjacent pairs of energy levels. Such revivals take place only for a non-zero detuning between the frequency of the atom and the cavity. This revival type is connected with  specific resonance arising between two adjacent pairs of energy levels. This type of resonance can be an obstacle in the non-linear optical quantum devices, which are used to construct quantum gates. In particular, the presence of a small detuning can lead to a fundamental limitation of the accuracy of quantum operations. However, such revival we can try to use also to improve the reliability of optical gates.
The set of parameters that provide non Rabi revival is dense in the space of all parameters; this set is obtained by solving the algebraic equations of the second order in rational numbers. Finding more complex types of revivals in this model is reduced to the solution in rational numbers of systems of nonlinear algebraic equations that has no algorithmic solution. Certain combinations of parameters ensure full revival for some pairs of not adjacent pairs of energy levels, for example, for some troikas of such pairs.

\section{Acknowledgements}

The work is partially supported by Russian Fond for Fundamental Researches, grant N 15-01-06132 a.


\begin{thebibliography}{99}
\bibitem{JC}E.T. Jaynes, F.W. Cummings, Comparison of quantum and semiclassical radiation theories with application to the beam maser, Proc. IEEE 51 (1): 89–109, (1963). doi:10.1109/PROC.1963.1664
\bibitem{DF}J. P. Dowling, J. D. Franson, H. Lee, and G. J. Milburn, Towards Scalable Linear-Optical Quantum Computers, Quantum Information Processing, Vol. 3, Nos. 1–5, October 2004.
\bibitem{KLM}E. Knill, R. Laflamme, G. J. Milburn, A scheme for efficient quantum computation with linear optics, Nature 409, 46-52 (4 January 2001) | doi:10.1038/35051009.
\bibitem{KM}Pieter Kok, W.J. Munro, Kae Nemoto, T.C. Ralph, Jonathan P. Dowling, G.J. Milburn, Review article: Linear optical quantum computing, Rev. Mod. Phys. 79, 135 (2007).
\bibitem{Ka}Karatsuba A., Karatsuba E.,  A resummation formula for collapse and revival in the Jaynes-Cummings model,  Journal of Physics A: Mathematical and Theoretical, Volume 42, Issue 19, article id. 195304, 16 pp. (2009). 
\bibitem{Ma}Yuri Matiyasevich Hilbert's 10th Problem, The MIT Press, 1993. ISBN 0-262-13295-8.
\bibitem{HA}Himadri Shekhar Dhar, Arpita Chatterjee, Rupamanjari Ghosh, Mapping generalized Jaynes-Cummings interaction into correlated finite-sized systems, J. Phys. B: At. Mol. Opt. Phys. 47, 135501 (2014).
\bibitem{HZ}Shu He, Yang Zhao, Qing-Hu Chen, Absence of collapse in quantum Rabi oscillations, Phys. Rev. A 90, 053848 (2014).
\bibitem{RH}E. Rasanen, E. J. Heller, Optimal control of quantum revival, Eur. Phys. J. B 86, 17 (2013). 
\bibitem{SF}Shane Dooley, Francis McCrossan, Derek Harland, Mark J. Everitt, Timothy P. Spiller, Collapse and Revival and Cat States with an N Spin System, Phys. Rev. A 87, 052323 (2013).
\bibitem{Sho}B.W.Shore, P.L.Knight, The Jaynes-Cummings Model, Journal of Modern Optics, vol. 40, issue 7, (1993), pp. 1195-1238. 
\bibitem{ZQ}Yu-Yu Zhang, Qing-Hu Chen, Shi-Yao Zhu, Jaynes-Cummings model:Counter rotating effect on the vacuum Rabi splitting and atom-cavity dynamics, Chinese Physics Letters 30, 114203 (2013).
\end{thebibliography}
\end{document}